\documentclass[final,5p,times,twocolumn,authoryear]{elsarticle}
\usepackage{amsmath}
\usepackage{amssymb}
\usepackage{amsthm}
\journal{Advances in Space Research}
\begin{document}
\begin{frontmatter}
\title{Nucleus-acoustic envelope solitons and their modulational instability in a degenerate quantum plasma system}
\author{*N. A. Chowdhury, M. M. Hasan, A. Mannan, and A. A. Mamun}
\address{Department of Physics, Jahangirnagar University, Savar, Dhaka-1342, Bangladesh.

*Email: nurealam1743phy@gmail.com}
\begin{abstract}
  The basic features of nucleus-acoustic (NA) envelope bright and dark solitons, which exist in degenerate quantum plasmas, have been
  theoretically investigated by deriving the nonlinear Schr\"{o}dinger (NLS) equation. The reductive perturbation method,
  which is valid for a small but finite amplitude limit, is employed. It is found that the bright envelope solitons are
  modulationally unstable, whereas the dark ones are stable. It is also observed that the fundamental properties (viz. Modulational instability (MI) growth
  rate, width and energy concentration of NA waves, etc.) of NA unstable bright envelope solitons are significantly
  modified by constituent particles number density. The implications of our results obtained from our present investigation
  in astrophysical compact objects like white dwarfs and neutron stars are briefly discussed.
\end{abstract}

\begin{keyword}
Modulational instability, Growth rate, Envelope solitons, degenerate, quantum.
\end{keyword}
\end{frontmatter}
\section{Introduction}
\label{}
%%%%%%%%%%%%%%%%%%%%%%%%%%%%%%%%%%%%%%%%%%%%%%%%%%%%%%%%%%%%%%%%%%
There is a large number of astrophysical compact objects (like white
dwarfs, neutron stars, black holes, etc.) which are formed under extreme conditions
\citep{Chandrasekhar1931,Chandrasekhar1935,Koester1990,Shapiro1983,Garcia-Berro2010}.
The particle number density of these compact objects is so high (order of $10^{30}~cm^{-3}$
in white dwarfs, and order of $10^{36}~cm^{-3}$ even more in neutron stars) that
the de Broglie wavelength of particles is comparable to the inter-particle
distance \citep{Koester1990,Shapiro1983}. Thus, the degenerate pressure and quantum effects, which play
an important role in understanding the dynamics of the constituents charged particles;
viz. electrons, positrons, and heavy nuclei \citep{Bedwehy2011,El-Labany2015,Hasan2016} of these compact objects,
need to be considered. Chandrasekhar \citep{Chandrasekhar1931} first assumed that hydrogen/helium nuclei and degenerate electrons are the main constituents of
compact objects like white dwarfs. On the other hand, Koester \citep{Koester1990} noticed that instead of hydrogen/helium nuclei, white dwarfs unusually
contain carbon/oxygen. The average density of particles in the core of white dwarfs is typically high ($\sim 10^{30} cm^{-3}$ ) compared
with outer mantle ($\sim 10^{26} cm^{-3}$ ), due to this, the particles are relativistically degenerate in the inner
core of white dwarfs, but are non-relativistically degenerate in the outer mantle \citep{Shukla2011}.
Recently, Payload for Antimatter Matter Exploration and Light-nuclei Astrophysics (PAMELA) satellite has measured the cosmic-ray positron
fraction \citep{Adriani2009,Adriani2010}. Kashiyama and Ioka \citep{Kashiyama2010} suggested that white dwarf pulsars
can compete with neutron star pulsars for producing the excesses of cosmic ray positrons and electrons ($e^+$ and $e^-$).
This means that in addition to degenerate electrons and C or O nuclei, white dwarfs contain
degenerate positrons \citep{Adriani2009,Adriani2010,Kashiyama2010}.
A number of works \citep{Hasan2016,Masood2011,Rahman2013,Hossen2014,Moghanjoughi2010,El-Taibany2012a} have been done by considering
the degenerate electron-positron-ion plasma. Mehdipoor and Esfandyari-Kalejahi \citep{Mehdipoor2012} studied
the ion acoustic waves in a plasmas system consisting of degenerate electrons, positrons, and isothermal ions.
They found that higher order corrections significantly change the properties of the Korteweg-de Vries solitons, and that
under certain conditions, both compressive and rarefactive solitary waves exist in such a plasma system.
Chandra and Ghosh \citep{Chandra2012} examined the MI of electron-acoustic waves by using
the quantum hydrodynamic model. They observed that the relativistic and degeneracy parameters modify the stability condition,
and properties of the envelope solitons. El-Taibany \citep{El-Taibany2012b} have investigated the
nonlinear propagation of fast and slow magnetosonic perturbation modes in non-relativistic, ultra-cold, degenerate
electron-positron plasma. They found that the basic features of the electromagnetic solitary structure are significantly
modified by the effects of degenerate electron and positron pressures. Recently, in a cold degenerate plasma, Mamun {\it et al.} \citep{Mamun2016}
introduced a concept of new electro-acoustic (NA) mode in which the inertia is provided by the nucleus mass density, and the restoring force is
provided by the degenerate non-inertial particle species pressure which depends only on particle number density. It is a new mode since it disappears
if we neglect the effect of electron or positron degeneracy. It should be noted here that ion/electron/positron-acoustic waves do not exist in cold plasma
limit, but the new NA waves exists in cold plasma limit. They studied the NA shock \citep{Mamun2016} and solitary \citep{Mamun2017}  structures associated
with this new mode. Ata-ur-Rahman \citep{Ata-ur-Rahman2015} studied MI of quantum-ion acoustic waves in a
plasma system consisting of inertial non-degenerate warm ions and inertialess relativistic degenerate electrons and positrons. Therefore, in our present work, we have examined the MI
of the NA waves (NAWs) propagating in such kinds of degenerate dense plasma (composed of inertial non-relativistically degenerate heavy nuclei,
inertialess ultra-relativistically degenerate electrons and positrons), which abundantly occur in astrophysical compact objects (e.g. white dwarfs, neutron stars).

The present paper is organized as follows. The basic governing equations of our plasma model are presented in Sec. \ref{Gov-Eq}.
Derivation of the NLS equation in Sec. \ref{Derivation}. The stability analysis and envelope solitons are
presented in Sec. \ref{Stability}. Discussion are presented in Sec. \ref{Discussion}.
\section{Governing Equations}
\label{Gov-Eq}
We have considered a plasma system consisting of inertial non-relativistic degenerate heavy nuclei, ultra-relativistically
degenerate electrons and positrons. At equilibrium, the quasi-neutrality condition
can be expressed as $Z_h n_{h0} + n_{p0}= n_{e0}$, where $n_{h0}$, $n_{p0}$, and $n_{e0}$ are
the equilibrium number densities of heavy nuclei, positrons, and electrons, respectively. The
normalized governing equations of the NAWs in our considered plasma system are given below
\begin{eqnarray}
&&\hspace*{-0.9cm}\frac{\partial n_h}{\partial t}+\frac{\partial}{\partial x}(n_h u_h)=0,\label{eq1}\\
&&\hspace*{-0.9cm}\frac{\partial u_h}{\partial t} + u_h\frac{\partial u_h }{\partial x}=-\frac{\partial \phi}{\partial x}-\beta\frac{\partial n^{2/3}_h}{\partial x},\label{eq2}\\
&&\hspace*{-0.9cm}\frac{\partial^2 \phi}{\partial x^2}=(1+\mu_p) n_e-n_h-\mu_p n_p.\label{eq3}
\end{eqnarray}
The pressure for heavy nuclei fluid \citep{Chandrasekhar1931} can be represented by the following equation:
\begin{eqnarray}
&&\hspace*{-0.9cm}P_h=K_h N^\alpha_h, \label{eq4}
\end{eqnarray}
where
\begin{eqnarray}
&&\hspace*{-0.9cm}\alpha=\frac{5}{3};~~ K_h=\frac{3}{5} \left(\frac{\pi}{3}\right)^\frac{1}{3}\frac{\pi \hbar^2}{m}\simeq\frac{3}{5} \Lambda_c \hbar c, \label{eq5}
\end{eqnarray}
\noindent for the nonrelativistic limit (where $\Lambda_c=\pi\hbar/m c=1.2\times {10}^{-10} cm$, and $\hbar$ is the
Planck constant $(h)$ divided by $2\pi$). While pressure for the electron and positron fluid are, respectively,
\begin{eqnarray}
&&\hspace*{-0.9cm}P_e=K_e N^\gamma_e,~~~P_p=K_p N^\gamma_p, \label{eq6}
\end{eqnarray}
and for the ultrarelativistic limit
\begin{eqnarray}
&&\hspace*{-0.9cm}\gamma=\frac{4}{3};~~ K_e=K_p=\frac{3}{4}\left(\frac{\pi^2}{9}\right)^\frac{1}{3} \hbar c \simeq\frac{3}{4} \hbar c.\label{eq7}
\end{eqnarray}
\noindent For inertialess ultra-relativistically degenerate electron and positron, we can obtain
the expressions for electron and positron number densities as, respectively,
\begin{eqnarray}
&&\hspace*{-0.9cm}n_e=(1+\phi)^3, \label{eq8}\\
&&\hspace*{-0.9cm}n_p=(1-\mu_e\phi )^3. \label{eq9}
\end{eqnarray}
Substituting Eqs. (\ref{eq8}) and (\ref{eq9}) into Eq. (\ref{eq3}), and then expanding up to third order, we get
\begin{eqnarray}
&&\hspace*{-0.9cm}\frac{\partial^2 \phi}{\partial x^2}=1-n_h+\gamma_1 \phi+\gamma_2 \phi^2+\gamma_3 \phi^3+\cdot\cdot\cdot\cdot, \label{eq10}
\end{eqnarray}
here
\begin{eqnarray}
&&\hspace*{-0.9cm}\gamma_1=3+3\mu_p(1+\mu_e),~~~\gamma_2=3+3\mu_p(1-\mu^2_e),\nonumber\\
&&\hspace*{-0.9cm}\gamma_3=1+\mu_p(1+\mu^3_e),  \nonumber\
\end{eqnarray}
and

~~$\beta=\frac{\Lambda_c n^{2/3}_{h0}}{2Z_h n^{1/3}_{e0}}$,~~~~~~~~$\mu_p= \frac{n_{p0}}{Z_h n_{h0}}$,~~~~$\mu_e= (\frac{n_{e0}}{n_{p0}})^{1/3}$,\\

~~$\beta_1=\frac{2\beta}{3}$,~~~~~~~~~~~~~$\beta_2=\frac{\beta}{9}$,~~~~~~~~~$\beta_3=\frac{4\beta}{81}$.\\
\noindent In Eqs. $(1)-(3)$, $(8)$, and $(9)$, $n_j$ is normalized by $n_{j0}$ (j = h, e, p; h for heavy nuclei, e for electrons, p for positrons);
$u_h$ is the heavy nuclei fluid speed normalized by the NAWs speed $C_h=(3\hbar c Z_h n^{1/3}_{e0} /m_h)^{1/2}$ (with $m_e$ ($m_h$) being
the electron (heavy nuclei) rest mass, $c$ being the speed of light in vacuum, and $Z_h$ is the charge state of heavy nuclei); $\phi$ is the
electrostatic wave potential normalized by $3\hbar c n^{1/3}_{e0}/e$ (with $e$ being the magnitude of an electron charge).
The time and space variables are normalized by ${\omega^{-1}_{ph}}=(m_h/4\pi Z^2_h e^2 n_{h0})^{1/2}$
and $\lambda_{Dh}=(3\hbar c n^{1/3}_{e0}/4 \pi Z_h e^2  n_{h0})^{1/2}$, respectively.
%%%%%%%%%%%%%%%%%%%%%%%%%%%%%%%%%%%%%%%%%%%%%%%%%%%%%%%%%%%%%%%%%%%%%%%%%%%%%%%%%%%%%%%%%
%%%%%%%%%%%%%%%%%%%%%%%%%%%%%%%%%%%%%%%%%%%%%%%%%%%%%%%%%%%%%%%%%%%%%%%%%%%%%%%%%%%%%%%
\section{Derivation of the NLS equation}
\label{Derivation}
To study the MI of the NAWs, we will derive the NLS equation by
employing the reductive perturbation method. So we first introduce
the independent variables are stretched as
\begin{eqnarray}
&&\hspace*{-0.9cm}\xi={\epsilon}(x  - v_gt),~~~~\tau={\epsilon}^2 t, \label{eq11}
\end{eqnarray}
where $v_g$ is the envelope group velocity to be determined later and $\epsilon ~(0<\epsilon<1)$ is a
small (real) parameter. Then we can
write a general expression for the dependent variables as
\begin{eqnarray}
&&\hspace*{-0.9cm}M(x,t)=M_0 +\sum_{m=1}^{\infty}\epsilon^{(m)}\sum_{l=-\infty}^{\infty}M_{l}^{(m)}(\xi,\tau)~\mbox{exp}[i l(kx-\omega t)], \nonumber\\
&&\hspace*{-0.9cm}M_l^{(m)}=[n_{hl}^{(m)}, u_{hl}^{(m)}, \phi_l^{(m)}]^T,~~~~ M_l^{(0)}=[1, 0, 0]^T, \label{eq12}
\end{eqnarray}
where k and $\omega$ are real variables representing the fundamental (carrier) wave number
and frequency respectively. $M_l^{(m)}$ satisfies the pragmatic condition $M_l^{(m)}= M_{-l}^{(m)^*}$, where the asterisk denotes
the complex conjugate. The derivative operators in the above equations are treated as follows:
\begin{eqnarray}
&&\hspace*{-0.9cm}\frac{\partial}{\partial t}\rightarrow\frac{\partial}{\partial t}-\epsilon v_g \frac{\partial}{\partial\xi}+\epsilon^2\frac{\partial}{\partial\tau},~~~~~\frac{\partial}{\partial x}\rightarrow\frac{\partial}{\partial x}+\epsilon\frac{\partial}{\partial\xi}. \label{eq13}
\end{eqnarray}
Substituting Eqs. (\ref{eq12}) and (\ref{eq13}) into Eqs. (\ref{eq1}), (\ref{eq2}), and (\ref{eq10}) and collecting power terms of $\epsilon$,
first-order $(m=1)$ equations with $l=1$
\begin{eqnarray}
&&\hspace*{-0.9cm}-i\omega n_1^{(1)}+iku_1^{(1)}=0,~~-i\omega u_1^{(1)}+ik\phi_1^{(1)}+ik\beta_1 n_1^{(1)}=0,\nonumber\\
&&\hspace*{-0.9cm} n_1^{(1)}-k^2\phi_1^{(1)}-\gamma_1 \phi_1^{(1)}=0.\label{eq14}
\end{eqnarray}
The solution for the first harmonics read as
\begin{eqnarray}
&&\hspace*{-0.9cm} n_1^{(1)}=\frac{k^2}{S}\phi_1^{(1)},~~~~~~~u_1^{(1)}=\frac{k \omega}{S}\phi_1^{(1)},\label{eq15}
\end{eqnarray}
where $S=\omega^2-\beta_1 k^2$. We thus obtain the dispersion relation for NAWs
\begin{eqnarray}
&&\hspace*{-0.9cm} \omega^2=\frac{k^2}{(k^2+\gamma_1)}+\beta_1 k^2.\label{eq16}
\end{eqnarray}
The second-order when $(m=2)$ reduced equations with $(l=1)$ are
\begin{eqnarray}
&&\hspace*{-0.9cm}n_1^{(2)}=\frac{k^2}{S}\phi_1^{(2)}+\frac{2ik\omega(v_g k-\omega)}{S^2} \frac{\partial \phi_1^{(1)}}{\partial\xi},\nonumber\\
&&\hspace*{-0.9cm}u_1^{(2)}=\frac{k \omega}{ S}\phi_1^{(2)} +\frac{i(\omega^2 +\beta_1 k^2)(v_g k-\omega)}{S^2} \frac{\partial \phi_1^{(1)}}{\partial\xi}, \label{eq16}
\end{eqnarray}
with the compatibility condition
\begin{eqnarray}
&&\hspace*{-0.9cm}v_g=\frac{\partial \omega}{\partial k}=\frac{\omega^2-(\omega^2-\beta_1 k^2)^2}{k\omega}.\label{eq18}
\end{eqnarray}
The second-harmonic mode of the carrier, which comes from nonlinear self interaction and it's arising
from the components $(l=2)$ for the second order $(m=2)$ reduced equations
\begin{eqnarray}
&&\hspace*{-0.9cm}n_2^{(2)}=C_1|\phi_1^{(1)}|^2,~~u_2^{(2)}=C_2|\phi_1^{(1)}|^2,~~\phi_2^{(2)}=C_3|\phi_1^{(1)}|^2,\label{eq19}
\end{eqnarray}
where
\begin{eqnarray}
&&\hspace*{-0.9cm}C_1=\frac{3\omega^2 k^4 +2C_3 S^2 k^2}{2S^3},~~~~~C_2=\frac{\omega C_1 S^2-\omega k^4 }{k S^2},\nonumber\\
&&\hspace*{-0.9cm}C_3=\frac{3\omega^2 k^4-2\gamma_2 S^3}{2S^3{(4k^2+\gamma_1)-2S^2 k^2}}. \nonumber\
\end{eqnarray}
Now, we consider the expression for  $(m=3, l=0)$ and $(m=2, l=0)$, which leads the zeroth harmonic modes. Thus we obtain
\begin{eqnarray}
&&\hspace*{-0.9cm}n_0^{(2)}=C_4|\phi_1^{(1)}|^2,~~u_0^{(2)}=C_5|\phi_1^{(1)}|^2,~~\phi_0^{(2)}=C_6|\phi_1^{(1)}|^2,\label{eq20}
\end{eqnarray}
where
\begin{eqnarray}
&&\hspace*{-0.9cm}C_4=\frac{2v_g \omega k^3+ \omega^2 k^2-\beta_2 k^4+C_6 S^2}{S^2(v^2_g-\beta_1)},\nonumber\\
&&\hspace*{-0.9cm}C_5=\frac{ v_g C_4 S^2-2\omega k^3 }{S^2},\nonumber\\
&&\hspace*{-0.9cm}C_6=\frac{2v_g \omega k^3+k^2\omega^2 -\beta_2 k^4-2\gamma_2 S^2(v^2_g-\beta_1)}{\gamma_1 S^2(v^2_g-\beta_1)-S^2}.\nonumber\
\end{eqnarray}
Finally, the third harmonic modes $(m=3)$ and $(l=1)$ and  with the help of Eqs. $(15)-(20)$,
 give a system of equations, which can be reduced to the following  NLS equation:
\begin{eqnarray}
&&\hspace*{-0.9cm}i\frac{\partial \Phi}{\partial \tau}+P\frac{\partial^2 \Phi}{\partial \xi^2}+Q|\Phi|^2\Phi=0, \label{eq21}
\end{eqnarray}
where $\Phi=\phi_1^{(1)}$ for simplicity. The dispersion coefficient $P$ is
\begin{eqnarray}
&&\hspace*{-0.9cm}P=\frac{v_g\beta^2_1 k^5 -3 v_g k \omega^4+4\beta_1 k^2 \omega^3 +2v_g\beta_1 \omega^2  k^3  -4\omega \beta^2_1 k^4 }{2\omega^2 k^2},\nonumber\
\end{eqnarray}
and the nonlinear coefficient $Q$ is
\begin{eqnarray}
&&\hspace*{-0.9cm} Q=\frac{S^2}{2\omega k^2}\left[2\gamma_2(C_3+C_6)+3\gamma_3-\frac{\omega^2k^2(C_1+C_4)}{S^2} \right.\nonumber\\
&&\hspace*{3.2cm}\left.-\frac{2\omega k^3(C_2+C_5)}{S^2}-\frac{\beta_3 k^8}{S^4}\right].\nonumber\
\end{eqnarray}
%%%%%%%%%%%%%%%%%%%%%%%%%%%%%%%%%%%%%%%%%%%%%%%%%%%%%%%%%%%%%%%%%%%%%%%%
\begin{figure*}[htp]
  \centering
  \begin{tabular}{ccc}
  \includegraphics[width=70mm]{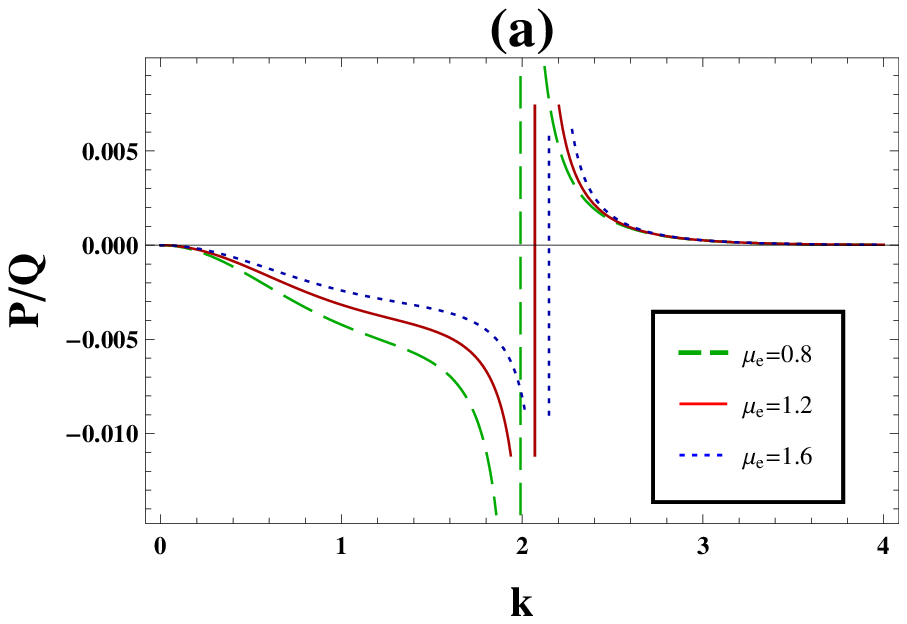}&
  \hspace{0.15in}
  \includegraphics[width=70mm]{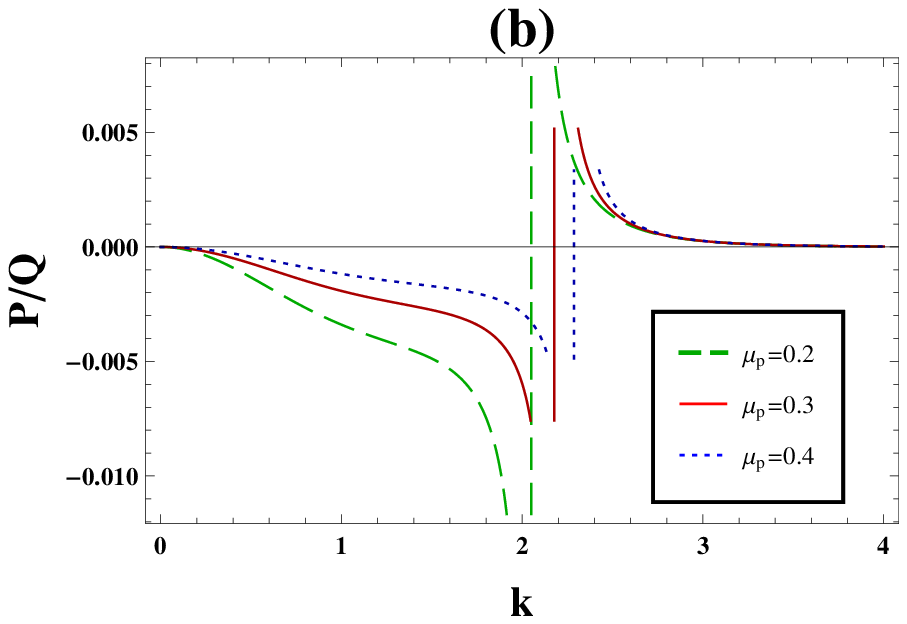}\\
  \includegraphics[width=70mm]{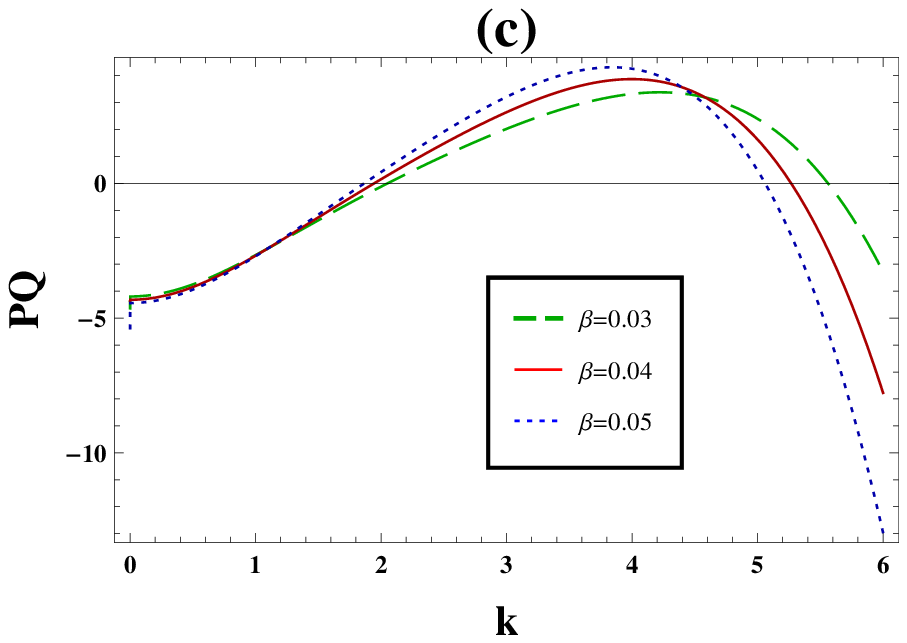}&
  \hspace{0.15in}
  \includegraphics[width=70mm]{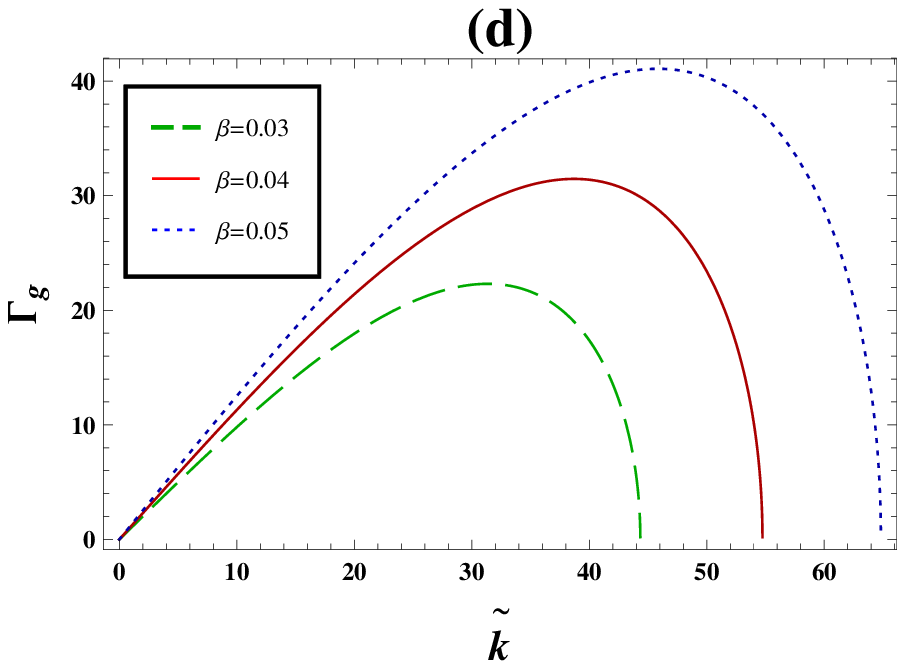}\\
  \end{tabular}
  \label{figur}\caption{Showing the variation of $P/Q$ or $PQ$ against $k$ for different values of plasma parameters,
  (a) $P/Q$ vs $k$ for $\mu_e$, (b) $P/Q$ vs $k$ for $\mu_p$, and (c) $PQ$ vs $k$ for $\beta$. (d) Plot of the of MI growth
  rate $(\Gamma_g)$ against $\tilde{k}$ for $\beta$, along with $k=3$ and $\Phi_0=0.5$. Generally, all the figures are generated by using
  these values $\beta=0.03$,\,$\mu_e=1.1$, and $\mu_p=0.2$. Where $Z=12,\, n_{e0}=9.3\times{10}^{29},\,n_{p0}=0.7n_{e0}$, and $n_{h0}=0.5n_{e0}$.}
\end{figure*}
\begin{figure*}[htp]
  \centering
  \begin{tabular}{ccc}
  \includegraphics[width=70mm]{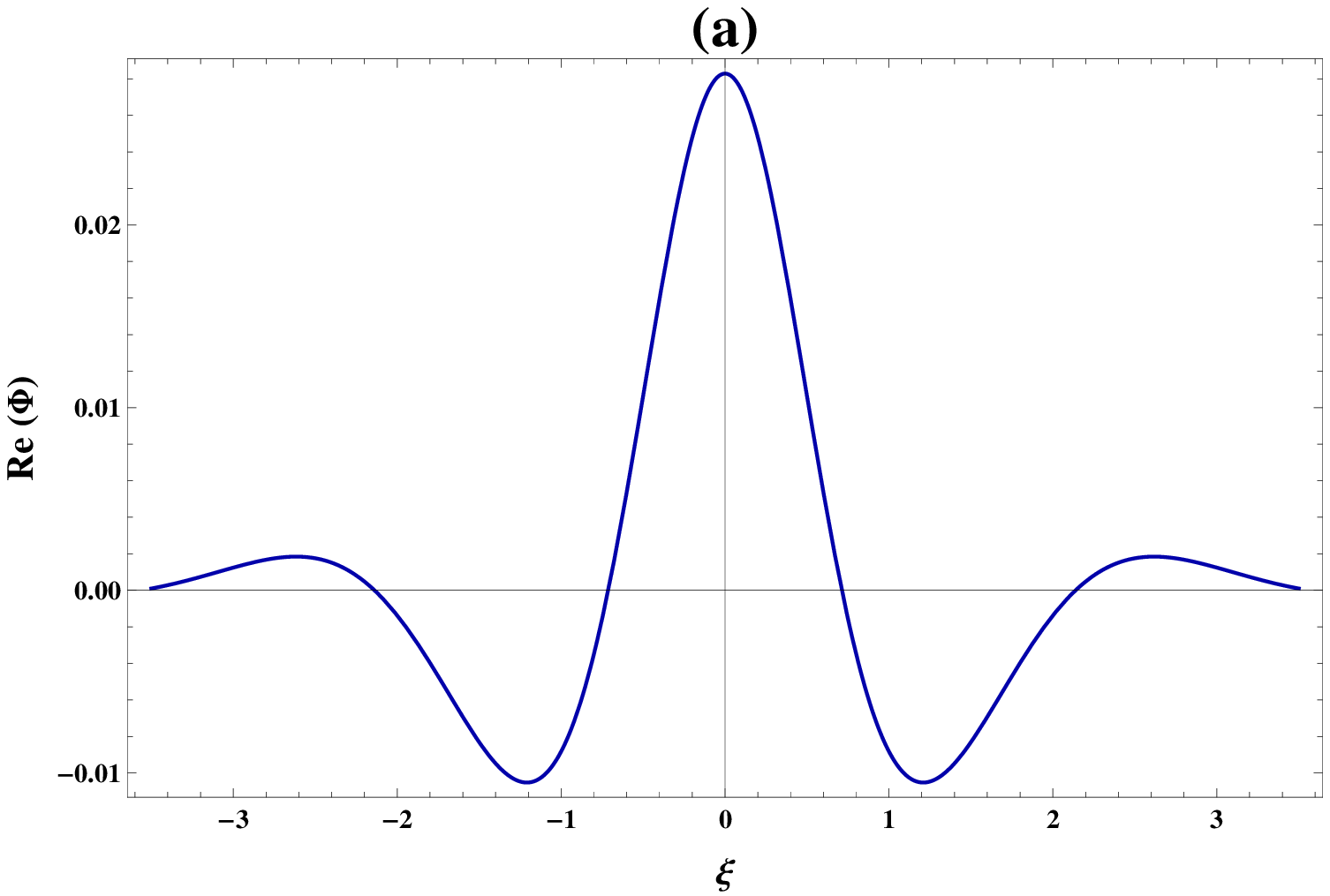}&
  \hspace{0.2in}
  \includegraphics[width=70mm]{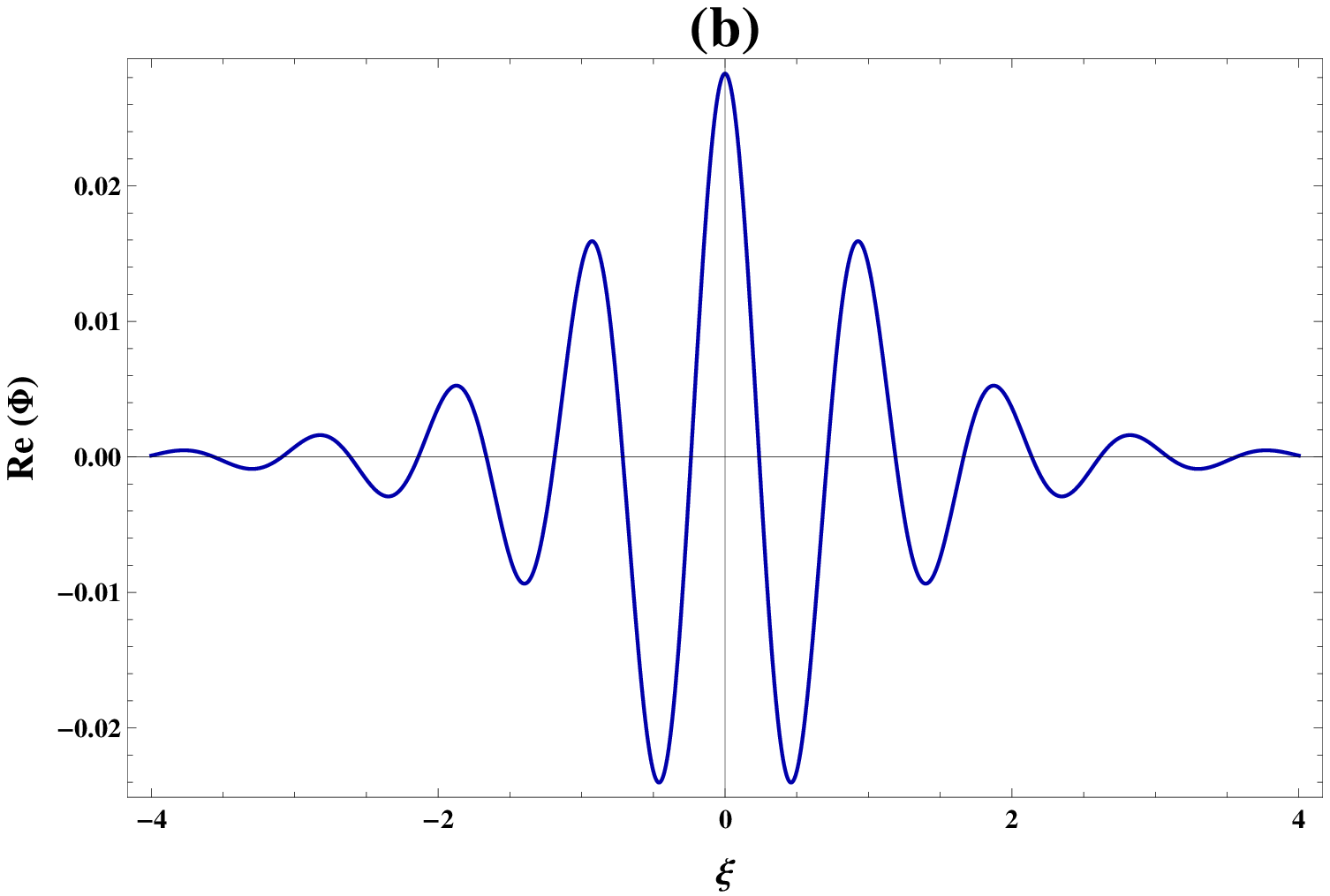}&\\
  \includegraphics[width=70mm]{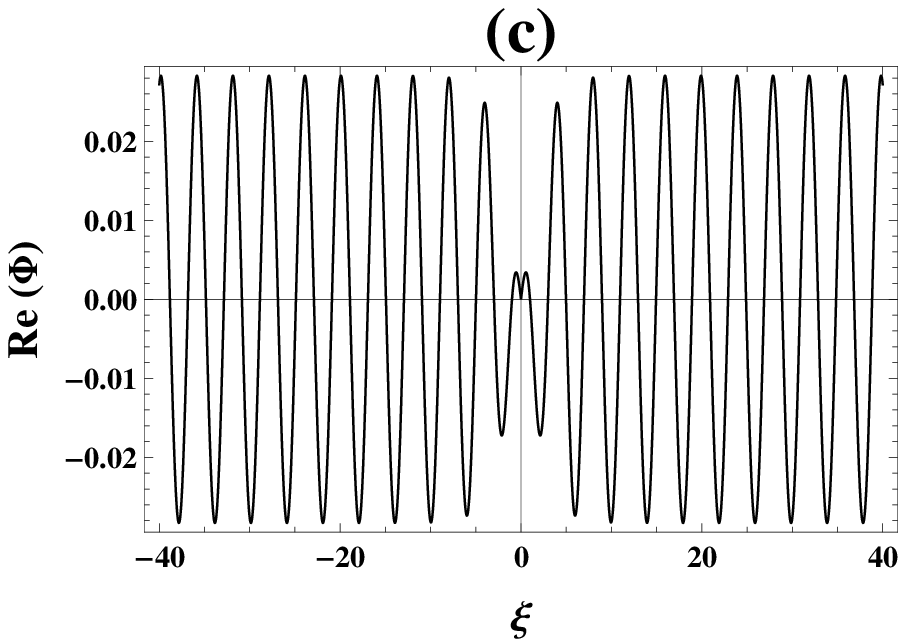}&
  \hspace{0.2in}
  \includegraphics[width=70mm]{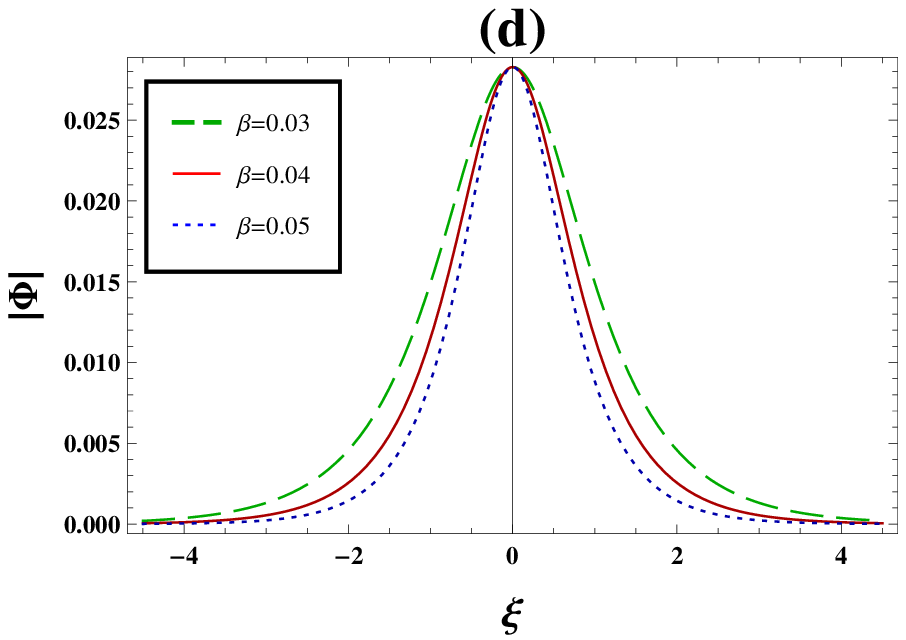}\\
  \end{tabular}
  \label{figur}\caption{Showing the variation of $Re(\Phi)$ against $\xi$,
  (a) Real part of bright envelope soliton for $k=3$ and $U=0.1$, (b) Real part of bright envelope solitons for $k=3$ and $U=0.3$,
  (c) Real part of dark envelope solitons for $k=1$ and $V=0.3$. (d) The variation of the $|\Phi|$ of bright envelope solitons against $\xi$,
  for $\beta$, and $k=3$. Along with $\tau=0$, $\psi_{b0}=\psi_{d0}=0.0008$, and $\Omega_0=0.4$.}
\end{figure*}
%%%%%%%%%%%%%%%%%%%%%%%%%%%%%%%%%%%%%%%%%%%%%%%%%%%%%%%%%%%%%%%%%%%%%%%%%%
\section{Stability analysis and envelope solitons}
\label{Stability}
The analysis of the NLS equation reveals the existence of either stable or unstable pulses.
The expedient tool to examine the region of stability/instability of the amplitude of envelope pulse
for external perturbations is the sign of the ratio $P/Q$. When the sign of the ratio $P/Q$ is negative,
the modulated envelope pulse is stable and dark envelope solitons exist, while the sign of the
ratio $P/Q$ is positive, the modulated envelope will be unstable against external perturbations
and in this region bright envelope solitons exist. For different values of plasma parameters,
the variation of the ratio $P/Q$ with carrier wave number $k$ is depicted in Figs. $1(a)$ and  $1(b)$.
In all cases studied here, $Q=0$ corresponds to zero dispersion point leading to $P/Q\rightarrow\pm\infty$.
The corresponding value of $k(=k_c)$ is called critical or threshold wave number for the onset of MI. It
can be shown from Fig. $1(a)$ that with the increase of electron number density (via $\mu_e$), the value
of $k_c$ increases. Excess number electrons are provided more restoring force which may be extended the
stability domain of NAWs. Similar behaviour is also observed in Fig. $1(b)$ for the positron number density
(via $\mu_p$). This means that the $k_c$  value is so much sensitive to change in the inertialess
constituent species number density.

Now, by considering a harmonic wave solution of Eq. (\ref{eq21}), of the form $\Phi=\Phi_0 e^{iQ{|\Phi_0|^2}\tau}$, where $\Phi_0$
denotes the constant amplitude of the carrier wave. One immediately obtains the nonlinear dispersion relation
for the amplitude modulation of NAWs packets:
\begin{eqnarray}
&&\hspace*{-0.9cm}\tilde{\omega}^2=P^2{\tilde{k}}^2 \left({\tilde{k}}^2-2\frac{Q}{P}{|\Phi_0|^2}\right),\label{eq22}
\end{eqnarray}
where $\tilde{k}$ and $\tilde{\omega}$ are the perturbed wave number and the frequency, respectively.
Clearly, if $PQ<0$, for all values of $\tilde{k}$, the NAWs are stable in the presence of small perturbation since
$\tilde{\omega}$ is always real. On the other hand, when $PQ>0$, the MI would set in as $\tilde{\omega}$ becomes imaginary, and
the envelope pulse is unstable for $\tilde{k}<k_c=\sqrt{2Q{|\Phi_0|}^2/P}$. In the region $PQ>0$ and $\tilde{k}<k_c$, the
growth rate ($\Gamma_g$) of MI is given by
\begin{eqnarray}
&&\hspace*{-0.9cm}\Gamma_g=|P|~{\tilde{k}^2}\sqrt{\frac{{k^2_{c}}}{\tilde{k}^2}-1}. \label{eq23}
\end{eqnarray}
Clearly, the maximum value $\Gamma_{g(max)}$ of $\Gamma_g$ is obtained at $\tilde{k}=k_c/\sqrt{2}$
and is given by $\Gamma_{g(max)}=|Q||\Phi_0|^2$. The effects of heavy nuclei number density (via $\beta$) on growth rate of the
MI is depicted in Fig. $1(d)$. It is observed that the growth rate is significantly
affected by variation of $\beta$. The growth rate of MI increases with
the increase of heavy nuclei number density. As the value of heavy nuclei number increases it begins to contribute large
to the moment of inertia, then the value of the MI growth rate increases rapidly. Moreover,
the growth rate increases with increases of $\tilde{\kappa}$. For a particular value of $\tilde{\kappa}$, the
growth rate is obtained it's critical value $(\Gamma_g\equiv\Gamma_{gc})$. Hence further increase the values
of $\tilde{\kappa}$ make the growth rate  decreases significantly.
\subsection{Bright envelope solitons}
For the same sign of $P$ and $Q$ (for $PQ>0$), i.e., for larger wave numbers [see Fig. $1(c)$],
the monochromatic waves are modulationally unstable, and this can lead to the formation of a bright
envelope solitons, localized envelope pulses of the form depicted in Figs. $2(a)$ and $2(b)$.
In this case, an exact analytical soliton solution of the Eq. (\ref{eq21}) can be obtained by considering
$\Phi=\sqrt{\psi}~\mbox{exp}(i\theta)$, where $\psi$ and $\theta$ are real functions as \citep{Kourakis2005,Shukla2002,Schamel2002,Fedele2002}
\begin{eqnarray}
&&\hspace*{-0.9cm}\psi=\psi_{b0}~ \mbox{sech}^2 \left(\frac{\xi-U\tau}{W_b}\right),\nonumber\\
&&\hspace*{-0.9cm}\theta=\frac{1}{2P} \left[U \xi+ \left(\Omega_0-\frac{U^2}{2}\right)\tau\right],\label{eq24}
\end{eqnarray}
here, $U$ is travelling speed of the localized pulse and oscillating at a frequency $\Omega_0$ (for $U=0$).
The relation between the pulse width $(W_b)$ and the constant maximum amplitude $\psi_{b0}$ is
\begin{eqnarray}
&&\hspace*{-0.9cm}W_b=\sqrt{\frac{2P}{Q \psi_{b0}}}.\label{eq25}
\end{eqnarray}
The effect of heavy nuclei number (via $\beta$) on the NAWs is shown in Fig. $2(d)$ and it can be seen that with the
increases $\beta$ the width of NAWs is also decreases, but the amplitude of the NAWs profile remains constant. This happens due
to the fact that when the heavy number density is increasing then the moment of inertia of the
plasma system is also increased which leads to formation of less energetic profile.
%%%%%%%%%%%%%%%%%%%%%%%%%%%%%%%%%%%%%%%%%%%%%%%%%%%%%%%%%%%%%%%%%%%%%%%%%%%%%%%%%%%%%%%%%%%%%%%%%%%%%%%%%%%%%%%%%%%%%%%%
\subsection{Dark envelope solitons}
For the opposite sign of $P$ and $Q$ (namely for $PQ<0$), the modulationally stable region is obtained. We find dark envelope solitons,
which general analytical form reads as
\begin{eqnarray}
&&\hspace*{-0.9cm}\psi=\psi_{d0}~ \mbox{tanh}^2 \left(\frac{\xi-V\tau}{W_d}\right),\nonumber\\
&&\hspace*{-0.9cm}\theta=\frac{1}{2P}\left[V \xi-\left( \frac{V^2}{2}-2PQ\psi_{d0}\right)\tau \right],\label{eq26}
\end{eqnarray}
here, $V$ is travelling speed of the localized region of hole (void) and oscillating at a frequency $\Omega_0$ (for $V=0$).
The relation between the pulse width $(W_d)$ and the constant maximum amplitude $\psi_{d0}$ is
\begin{eqnarray}
&&\hspace*{-0.9cm}W_d=\sqrt{\frac{2|P/Q|}{ \psi_{d0}}}.\label{eq27}
\end{eqnarray}
Dark envelope solitons [real part of Eq. (\ref{eq26})] is depicted in Fig. $2(c)$.

\section{Discussion }
\label{Discussion}
We have investigated the amplitude modulation of NAWs packet in an unmagnetized multi-component plasma system consisting
of inertial non-relativistically degenerate heavy nuclei, ultra-relativistically degenerate electrons, and positrons. By
employing the reductive perturbation method, a NLS equation is derived, which governs the evolution of NAWs.
Both stable and unstable regions are found. The results, we have found from this investigation can be summarized as follows:
\begin{enumerate}
\item{The $k_c$ value may recognize the stability of NAWs. For large $k(k>k_c)$,
  there is modulationally unstable region where bright envelope solitons exist, on the other hand for small $k(k<k_c)$,
  there is stable region (dark envelope solitons exist in this region) are  found. The $k_c$ value is shifted towards
  higher values as the increase of electrons/positron number density (via $\mu_e/\mu_p$), respectively. So inertialess species play
  a crucial role to change stability condition.}
\item{MI growth rate is significantly affected by non-relativistically degenerate heavy nuclei number density (via $\beta$). With the increase of
  non-relativistically degenerate heavy nuclei number density, the maximum value of the MI growth rate is also increased rapidly.
  So excess number of heavy nuclei enhanced the MI growth rate.}
\item{With the increasing of non-relativistically degenerate heavy nuclei number density (via $\beta$), the amplitude of the
  envelope soliton remains constant, but the width of the NAWs decreases. This happens due to the fact of excess heavy nuclei number
  density, provides excessive  moment of inertia which may squeeze the width of the NAWs and associated
  energy concentration through the wave profile.}
\end{enumerate}
We hope that our investigation will be helpful for understanding the nonlinear electrostatic structures that may be
formed in some astrophysical compact objects (such as white dwarfs and neutron stars) containing ultra-relativistic
degenerate electrons and positrons, along with inertial non-relativistic degenerate heavy nuclei.
We find up through our analysis, how a degenerate plasma system behaviour depends on their constituent elements number density, not on their temperature.

\section*{Acknowledgements}
N. A. Chowdhury is grateful to the Bangladesh Ministry of Science and Technology for
awarding the National Science and Technology (NST) Fellowship.




\end{document}